\documentclass[reprint,amsmath,amssymb,aps]{revtex4-1}
\usepackage[colorlinks,citecolor=blue,linkcolor=blue,anchorcolor=blue,filecolor=blue,
urlcolor=blue]{hyperref}
\usepackage{graphicx}
\usepackage{ulem}
\usepackage{dcolumn}
\usepackage{bm}
\usepackage{todonotes}

\begin{document}

\title{Tidal deformability of strange stars and the GW170817 event}

\author{Odilon Louren\c{c}o, C\'esar H. Lenzi, Mariana Dutra}
 \affiliation{Departamento de F\'isica, Instituto Tecnol\'ogico de Aeron\'autica, DCTA, 12228-900, S\~ao Jos\'e dos Campos, SP, Brazil}
 
\author{Efrain J. Ferrer, Vivian de la Incera}%
\affiliation{Dept. of Physics and Astronomy, University of Texas at Rio Grande Valley, Edinburg, TX 78539, USA}%

\author{Laura Paulucci}
\affiliation{Universidade Federal do ABC, Av dos Estados, 5001, Santo Andr\'e, SP, 09210-170, Brazil}%

\author{J. E. Horvath}
\affiliation{Instituto de Astronomia, Geof\'isica e Ci\^encias Atmosf\'ericas/Universidade de S\~ao Paulo, Rua do Mat\~ao 1226, 05508-900 S\~ao Paulo SP, Brazil}%

\date{\today}
             
\begin{abstract}
In this work we consider strange stars formed by quark matter in the color-flavor-locked (CFL) phase of color superconductivity. The CFL phase is described by 
a Nambu-Jona-Lasinio model with four-fermion vector and diquark interaction channels. The effect of the color superconducting medium on the gluons are incorporated into the model by including the gluon self-energy in the thermodynamic potential. We construct parametrizations of the model by varying the vector coupling $G_V$ and comparing the results to the data on tidal deformability from 
the GW170817 event, the observational data on maximum masses from massive pulsars such as the MSP J0740+6620, and the mass/radius fits to NICER data for PSR J003+0451. Our results points out to windows for the~$G_V$ parameter space of the model, with and without gluon effects included, that are compatible with all these astrophysical constraints, namely, $0.21<G_V/G_S<0.4$, and $0.02<G_V/G_S<0.1$, respectively. We also observe a strong correlation between the tidal deformabilites of the GW170817 event and $G_V$. Our results indicate that strange stars cannot be ruled out in collisions of compact binaries from the structural point of view. 
\end{abstract}

\maketitle

\section{\label{sec:intro}Introduction}

A large degree of interest in the community was prompted by the announcement of a gravitational signal identified as the merging of two neutron stars (NS's). The GW170817 event \cite{Abbott2017} triggered an alert followed by many observatories and satellites, and at least 70 positive detections were reported. Among the most important observational highlights, a gamma-ray burst (GRB) definitely associated with the event \cite{AbbottGRB} confirmed the expectation that ``short'' GRBs are produced by the mergers, although the referred event was particularly faint (probably due to off-axis emission \cite{Tsvi}) and its recognition has been disputed \cite{Istvan}. 

Important observations of the light-curve showed, on the other hand, a distinctive IR excess a few days after the outburst, of the type now known as ``kilonovae'' \cite{AbbottGRB, Valenti}. It was linked to the production of lanthanides and actinides \cite{Nucleo}, given that high-opacity in the ejecta neatly explains the temporal behavior \cite{Nucleo2}. The recent identification of strontium in the spectrum of the source \cite{Watson2019} added credibility to this interpretation. Actinides are also expected to form in the event, perhaps dominating the production of many heavy isotopes in the galaxy and populating the end of the Periodic Table \cite{Nucleo3}.

In spite of this benchmark advance, the theory of NS merging still needs to provide many answers for 
the whole picture to be complete and compelling. This is quite a difficult task and should involve a 
number of physical ingredients and high-performance computation. One key ingredient is the 
composition of the matter in colliding stars. Even within the standard picture, nucleons are hardly 
the only particle present, since hyperons are expected at certain inner density and explicitly 
considered for many years~\cite{ref1,ref2,ref3,ref4,ref5,ref6,ref7,Yo}. Even more exotic components 
have been considered, notably quark matter, both as a part of the innermost region of the stars or 
as an absolutely stable state ({\it strange quark matter}, SQM) composing essentially all the star 
up to the upper layers~\cite{Bodmer,Witten,Terazawa,Itoh}. The latter idea has been around for more 
than three decades, and some indirect observational evidence for its possible existence has been 
given in Refs.~\cite{SS-Evidences}.

To establish the link between the star internal composition and the observational data is the main goal of many investigations. In this regard, certain knowledge can be crucial for discriminating among the proposed compositions, as for example, the results of reliable calculations of the nucleosynthesis process \cite{nos}, the so-called {\it tidal deformability} that can be extracted from the GW170817 data~\cite{lvc1,lvc2,Abbott2}, the star mass-radius relationship, etc.

In this paper we investigate if the equation of state~(EOS) of a strange star, described by a color-flavor-locked~(CFL) model with vector interactions and the gluon self-energy contribution \cite{Our-Gluons}, satisfies several observational constraints derived from the tidal deformability  inferred from the GW170817 event, the maximum-mass constraints from various known pulsars, and mass-radius estimates derived from the Neutron Star Interior Composition Explorer (NICER) data. When considering the information from the GW170817 event, we will assume that the two stars participating in the binary NS coalescence have the same EOS.

We show that the deformability parameter space predicted by our model matches the one obtained from the GW170817 data \cite{Abbott2}.  Furthermore, the maximum-mass constraints corresponding to PSR J1614-2230, PSR J0348+0432, and MSP J0740+6620 with $M=1.97\pm 0.04M_{\odot}$~\cite{Demorest}, $M=2.01\pm 0.04M_{\odot}$~\cite{Antoniadis}, and $2.14{{+0.10}\atop{-0.09}} M_{\odot}$~\cite{Cromartie}, respectively, are satisfied for the parameter values under consideration. In this matter, the inclusion of gluon effects increases the range of $G_V$ compatible with the observations. The reason for this is that the combined effect of the vector interactions and the gluon contribution makes the strange matter malleable, so that it can be sufficiently deformed while stiff enough to reach a high maximum mass. We also verify that the calculated dimensionless tidal deformabilites are of the same order as those obtained in relativistic and non-relativistic hadronic models studied in Refs.~\cite{had2,had3,had4,had5}.  

This paper is organized as follows: In Sec.~\ref{sec:model}, we introduce the model and its EOS, which then is used as input to solve the Tolman-Oppenheimer-Volkoff (TOV) equations. The definitions of the tidal deformabilities are presented in Sec.~\ref{tidaldef}. Our results and comparisons with the GW170817 event are shown in Sec.~\ref{results}.  Finally, we present the summary and concluding remarks of our study in Sec.~\ref{conclusions}.

\section{\label{sec:model}Modelling of self-bound compact stars}

According to the Bodmer–Terazawa–Witten (BTW) hypothesis \cite{Bodmer, Terazawa, Witten}, strange matter, which consists of roughly equal numbers of up, down, and strange quarks at high densities, is conjectured to be absolutely stable (it has lower energy per baryon than ordinary iron nuclei). If this is the case, the whole interior of a NS will likely be converted into strange matter. 

On the other hand, the ground state of the superdense quark system is unstable with respect to the formation of diquark condensates \cite{CS}, a non-perturbative phenomenon essentially equivalent to the Cooper instability of BCS superconductivity. Given that in QCD one gluon exchange between two quarks is attractive in the color-antitriplet channel, at sufficiently high density and sufficiently small temperature quarks should condense into Cooper pairs, which are color antitriplets. At densities much higher than the masses of the $u$, $d$, and $s$ quarks (a condition usually written as $\mu\gtrsim m_s^2/2\Delta$, with $m_s$ being the strange quark mass and $\Delta$ the pairing gap), one can assume that the three quarks are massless. In this asymptotic region the most favored state is the CFL phase \cite{CFL}, characterized by a spin-zero diquark condensate antisymmetric in both color and flavor. 

When considering CFL matter through a Nambu-Jona-Lasinio (NJL) model with four-fermion interactions at finite density, other interactions, besides the diquark channel \cite{DiQuark}, can also be considered.  Among these additional channels of interactions, vector interactions \cite{Vector-Int,dynamical} are the most relevant as they can significantly affect the stiffness of the EOS, and hence they will be considered in our analysis. On the other hand, gluons degrees of freedom are usually disregarded as negligible at zero temperature and finite density. However, in the color superconducting background, the gluons acquire Debye ($m_D$) and Meissner ($m_M$) masses
\begin{eqnarray}\label{Masses}
m_D^2=\frac{21-8\ln 2}{18} m_g^2,\;\;\;\;\;\; m_M^2=\frac{21-8\ln 2}{54}m_g^2,\nonumber
\\
m_g^2=g^2\mu^2N_f/6\pi^2.  \qquad \qquad\qquad
\end{eqnarray}
that depend on the chemical potential $\mu$ \cite{Gluon-Mass} and thus can affect the EOS of the CFL phase \cite{Our-Gluons}. In (\ref{Masses}), $N_f$ is the number of flavors and $g$ is the quark-gluon gauge coupling constant. Then, the net effect of the gluons in the CFL background is a $\mu$-dependent contribution that increases the energy density and decreases the pressure. 

The CFL thermodynamic potential with the contributions of the vector interactions and the gluons takes the form 
\begin{equation}\label{modelo}
\Omega_{\mbox{\tiny CFLg}}=\Omega_{q}+ \Omega_{g}  - \Omega_{vac}
\end{equation}
where the quark contribution at zero temperature is
\begin{eqnarray} \label{Gamma0}
\Omega_{q}&=&-\frac{1}{4\pi^2}\int_0^{\Lambda_{\rm cut}} dp p^2 (16|
\varepsilon|+16|\overline{\varepsilon}|)\nonumber
\\
&-&\frac{1}{4\pi^2}\int_0^{\Lambda_{\rm cut}}
dp p^2 (2|\varepsilon'|+2|\overline{\varepsilon'}|) +\frac{3\Delta^2}{G_D}-G_V\rho^2,
\end{eqnarray} 
with
\begin{equation}\label{Spectra}
\varepsilon=\pm \sqrt{(p-\tilde{\mu})^2+\Delta^2}, \quad
\overline{\varepsilon}=\pm \sqrt{(p+\tilde{\mu})^2+\Delta^2},\nonumber
\end{equation}
\begin{eqnarray}\label{Spectra-2}
\varepsilon'=\pm \sqrt{(p-\tilde{\mu})^2+4\Delta^2,}\quad
 \overline{\varepsilon}'=\pm \sqrt{(p+\tilde{\mu})^2+4\Delta^2}.\qquad
\end{eqnarray}
with $\tilde{\mu}=\mu-2G_V \rho$, and
\begin{eqnarray}
 \Omega_{g} &=& \frac{2}{\pi^2}\int_0^{\Lambda_{\rm cut}} dp 
p^2\left[3\sqrt{p^2+\tilde{m}_M^2\theta(\Delta-p)}\right.
 \nonumber \\
 &+&\left.\sqrt{p^2+ \tilde{m}^2_D \theta({\Delta}-p)+3\tilde{m}^2_g\theta(\tilde{\mu}-p)\theta(p-\Delta)} \right]
 \label{TP-gluons-T0}
\end{eqnarray} 
 is the gluon contribution at $T=0$. In (\ref{modelo}) we subtracted the vacuum constant 
$\Omega_{vac}\equiv \Omega_{\mbox{\tiny CFLg}}(\mu=0, \Delta=0)$.

The dynamical quantities $\Delta$ and  
$\rho$ are found from the equations
\begin{equation} \label{Gap-Eq1}
\frac{\partial\Omega_{\mbox{\tiny CFLg}}}{\partial\Delta} = 0, \;\;\;\; \rho=-\frac{\partial\Omega_q}{\partial\tilde{\mu}}
\end{equation}

The solution of the gap equation (first equation in (\ref{Gap-Eq1})) is a minimum of the thermodynamic potential while the solution of the second equation is a maximum \cite{Vector-Int}, since it defines, as usual in statistics, the particle number density $\rho=\langle\bar{\psi}\gamma_0\psi\rangle$.

Having the thermodynamic potential (\ref{modelo}), we can write the EOS of the system as
\begin{equation}\label{Pressure}
P_{\mbox{\tiny CFLg}}= -(\Omega_{q}+ \Omega_{g}  - \Omega_{vac})+(B-B_0), 
\end{equation} 
\begin{equation}\label{Energy}
\epsilon_{\mbox{\tiny CFLg}} = \Omega_{q}+ \Omega_{g}  - \Omega_{vac} + \tilde{\mu} \rho-(B-B_0)
\end{equation}

Notice that the chemical potential that multiplies the particle number density in the energy density is $\tilde{\mu}$ instead of $\mu$. This result can be derived following the same calculations of Ref. \cite{Israel} to find the quantum-statistical average of the energy-momentum tensor component $\tau_{00}$. 

In (\ref{Pressure})-(\ref{Energy}), we added the bag constant $B$, which in the NJL model can be dynamically found in the mean-field approximation in terms of the chiral condensates that exist at low density \cite{Oertel}. The vacuum bag constant $B_0=B|_{\rho_u=\rho_d=\rho_s=0}$ is introduced to ensure that $\epsilon_{\mbox{\tiny CFLg}}=P_{\mbox{\tiny CFLg}}=0$ in vacuum. Using the results of \cite{Oertel}, one can readily see that for the parameter set under consideration, the vacuum bag constant takes the value $B_0=B|_{\rho_u=\rho_d=\rho_s=0}=57.3$ MeV/fm$^3$. Moreover, at the high densities where the CFL phase occurs, the chiral condensates are all zero, and consequently $B=0$ \cite{Our-Gluons}. 

The mass-radius relationship of the system can be obtained using the EOS and the Tolman-Oppenheimer-Volkoff (TOV) equations
\begin{eqnarray}
\frac{dm(r)}{dr}&=&4\pi r^2\epsilon(r) \label{TOV1}\\
\frac{dP (r)}{dr} &=& -\frac{\left[\epsilon(r) + P(r)\right]\left[m(r) + 4\pi r^3 P (r)\right]}{r^2f(r)}
\label{TOV2}
\end{eqnarray}
written in natural units where $c = G = 1$. Here, $f(r)=1-2m(r)/r$, and $m(R)=M$ is the mass of the star with radius $R$. Since there is no strong evidence in favor of high spins in the GW170817 data, we shall not refer to this case.
Using these equations one can show that for each $G_V$, the gluons tend to decrease the maximum star mass in about $20 \%$~\cite{Our-Gluons}. The effect is even bigger at lower values of $G_V$. Sequences including gluons do not reach $2M_{\odot}$ unless $G_V/G_S>0.2$~\cite{Our-Gluons}. 

In the following sections, we will add new constraints to the mix to determine the compatibility of the CFL model -with and without gluons- with new observations like updated maximum mass values, tidal deformability of strange stars, and the mass-radius estimates obtained from NICER.

\section{Tidal Deformability}
\label{tidaldef}

The tidal deformability is a dynamical property of matter subject to a tidal field. Close analogy with known phenomena can be easily recognized from nuclear physics, in which several modes related to the nuclear structure (dipole, giant resonance, etc.) can be measured when the nucleus is subjected to perturbation (obviously not tidal). The linear regime of tidal deformability is seen every day in ocean tides. In the context of neutron star collision, tidal deformability is an extreme non-linear regime version of what occurs in bulk matter.

On very general grounds, and irrespective of a Newtonian or relativistic approach, the tidal deformability $\lambda \equiv {Q_{ij}\over{\varepsilon_{ij}}}$ is defined by the quotient of the induced quadrupole $Q_{ij}$ to the tidal field $\varepsilon_{ij}$, dimensionally expected to scale as the fifth power of the star radius $R^{5}$. In fact, introducing the
{\it gravitational Love number} $k_{2}$, the precise relation is
\begin{equation}
    \lambda = {2\over{3}} k_{2} R^{5}.
\label{lambda}
\end{equation}

Direct calculations of a collection of equations of state yield $k_{2} \sim 0.2-0.3$. For a general purpose, the tidal deformability can be made dimensionless dividing it by the mass of the star $M$ to the fifth power, namely
\begin{equation}
    \Lambda = {2\over{3}} k_{2} {R^{5}\over{M^{5}}} \equiv {2\over{3}} k_{2} C^{-5}
    \label{diml}
\end{equation}
where $C \equiv {M\over{R}}$ is the compactness. Numerically it can be seen that $\Lambda$ can vary three orders of magnitude
from its value for $\sim 1 M_{\odot}$ stars to the maximum mass of the configuration for a fixed EOS (and not considering other effects such as rotation, dynamical response of the tidal fields and magnetic fields). This is why many
works have focused on this quantity, which is very sensitive to the stars' composition \cite{kata}. Thus, even if we shall refer to one event (GW170817) only, its observation will be potentially important for an evaluation of the state of stellar interiors.

In addition to this novel test of the EOS, known tests must be also enforced to select out a realistic form of the pressure and energy density, for a given composition. This type of approach has been attempted in connection to the heavy-ion data, that is, a reconstruction of the allowed zones inferred \cite{Recons}. And of course, ``static'' information on neutron stars concerning the degree of stiffness of the EOS, allowing at least $2.14{{+0.10}\atop{-0.09}} M_{\odot}$~\cite{Cromartie} for the maximum mass, and a relatively large radius $13.02{{+1.24}\atop{-1.06}}$~km obtained~\cite{Col} with the
emission fits to the NICER data for PSR J0030+0451, with a determined mass of $1.44{{+0.15}\atop{-0.14}} M_{\odot}$, should be considered.

To proceed we must make contact with the problem of two compact stars colliding, not necessarily of the same mass. In the inspiral final phase of a binary system, periodic gravitational waves (GW) are emitted with a phase that can be expressed in a post-Newtonian expansion in powers of $v/c$ (also expressed as $u = (\pi M f)^{1/3}$ with $f$ the gravitational wave frequency), yielding a ``tidal'' term $\propto -(39/2) \tilde{\Lambda} u^{10}$, at the lowest order. The coefficient $\tilde{\Lambda}$ is given by
\begin{eqnarray}
    {\tilde{\Lambda}} = {16\over{13}}{{(M_{1}+12M_{2})M_{1}^{4}\Lambda_{1} + (M_{2}+12M_{1}) M_{2}^{4}\Lambda_{2}} \over {(M_{1}+M_{2})^{5}}}\qquad
\label{tilde}
\end{eqnarray}
where $\Lambda_{1}$ and $\Lambda_{2}$ are the dimensionless tidal deformabilities of each star as defined above. This result was first obtained by Flannagan and Hinder \cite{FH} and serves to investigate the response of the stellar material to the tidal field, as stated below, being extracted directly from the observed waveform.

\section{Numerical results and the GW170817 event}
\label{results}

In this section, we investigate how well the model described in section \ref{sec:model} of a 
self-bound compact star with CFL matter satisfies the tidal-deformability constraints imposed by the 
GW170817 event, the most recently observed maximum-mass values, and the mass/radius fits to NICER 
data for PSR J003+0451. We will consider CFL matter with and without gluons and discuss the region 
of compatibility on each case.

The model parameters used in the numerical calculations are defined by following a standard 
procedure, with the energy cutoff $\Lambda_{\rm cut}=602.3$ MeV and the 
quark-antiquark coupling $G_S{\Lambda^2_{\rm cut}}=1.835$ 
adjusted to fit $f_\pi$, $m_\pi$, $m_K$ and $m_{\eta'}$ to their empirical values in the sharp cutoff regularization \cite {Rehberg}. Then, the diquark coupling $G_D$, that produces a gap $\Delta\simeq10$ MeV at $\mu=500$ MeV, is found to be $G_D=1.2 G_S$. A similar ratio $G_D/G_S$ was already considered  in \cite{GD-GS} to investigate the $M-R$ relationship in hybrid compact stars with color superconducting cores. Changing $\Lambda$ in a few percentage, while simultaneously modifying $G_D$ to produce the same value of $\Delta$, does not affect our qualitative results. As for the values of the vector coupling, it is known that if the vector channel is originated from a Fierz transformation of a local color current-current interaction, the resulting coupling strength is $G_{V}=0.5 G_S$. If instead, one starts from the molecular instanton liquid model or the PNJL model, the Fierz transformations give rise to much smaller values of $G_V$ \cite{GV-Vacuum}. Based on these considerations, $G_V$ is usually taken as a free parameter in the range $G_V=(0-0.5)G_S$. Here we adopt this same range for $G_V$.

In order to correctly describe a strange star from this model, we consider stellar matter composed by $u$, $d$ and~$s$ quarks, in which the equations of state used as input to solve the TOV equations are given by $\epsilon=\epsilon_{\mbox{\tiny CFLg}}$ and $P=P_{\mbox{\tiny CFLg}}$ from (\ref{Pressure})-(\ref{Energy}) for the case with gluons, and the same equations but with $\Omega_g=0$, in the case without gluons. Once the inputs are defined, the solution of the TOV Eqs.~(\ref{TOV1}) and~(\ref{TOV2}) is constrained by the following conditions at the neutron star center: $P(0) = P_c$ (central pressure), and $m(0) = 0$ (central mass). The mass of the star for each set of parameters, is obtained as the solution of the TOV equations at the point where the pressure vanishes, i.e., when it reaches the surface of the star.

\begin{figure}[!htb]
\centering
\includegraphics[scale=0.35]{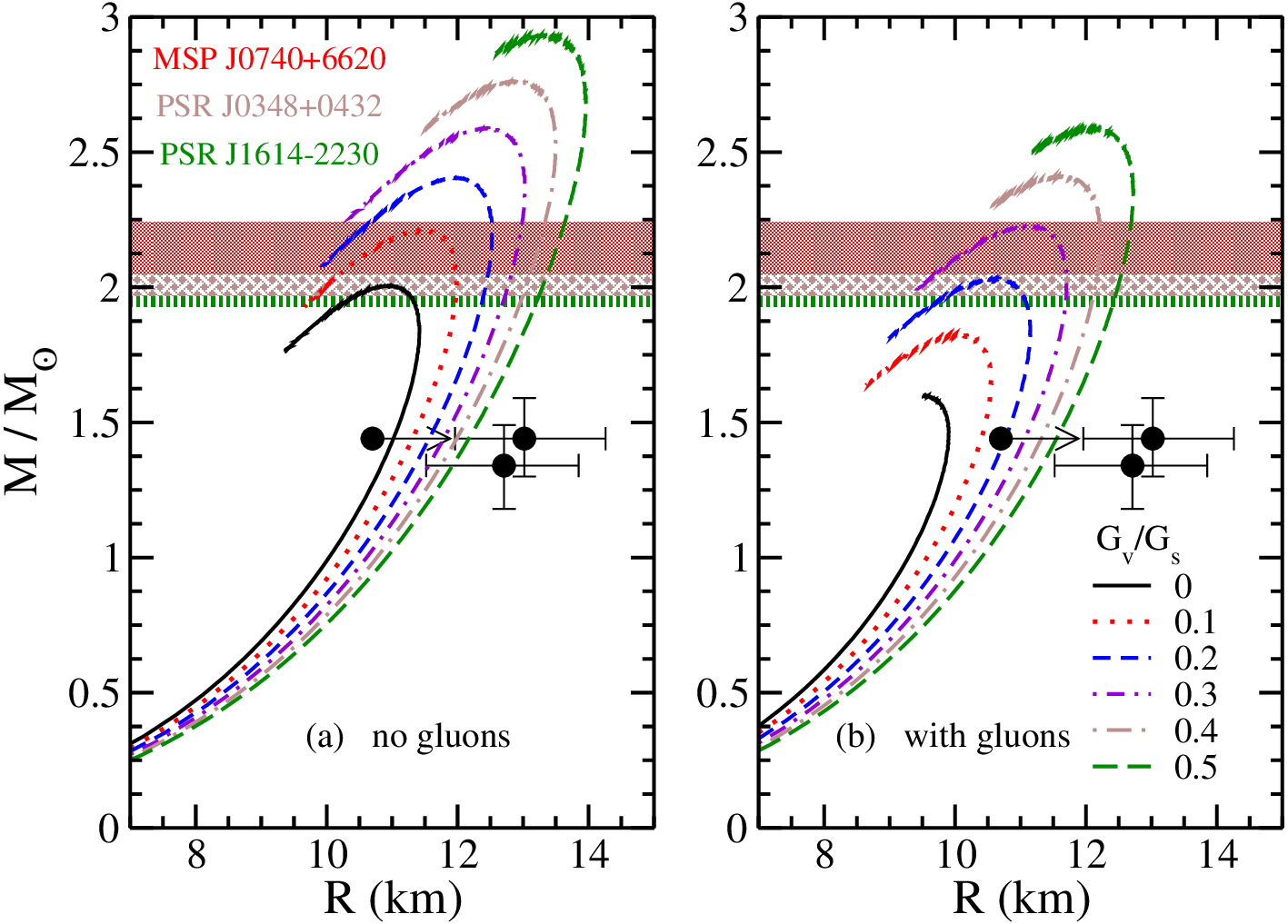}
\caption{Quark star mass, in units of $M_\odot$, as a function of its radius generated from the CFL phase (a) without and (b) with gluons contribution. Bands extracted from Refs.~\cite{Demorest,Antoniadis,Cromartie}. Circles with error bars are related to the NICER data~\cite{Col,l21,l25}.}
\label{mr}
\end{figure}

In Fig.~\ref{mr}, we present the mass-radius profiles of strange stars obtained with the NJL model used in this work, with and without the inclusion of gluons in its thermodynamics. The parametrizations were constructed by varying the vector channel strength of the model within the physically acceptable range of $G_V$.
From Fig.~\ref{mr}, it can be seen that the gluons' contribution reduces the value of the maximum star mass obtained in each parametrization. This result coincides with the one already reported in Ref.~\cite{Our-Gluons}. Here we go further in the analysis of these diagrams by comparing them with more recent observational data. Two of them are related to the mass values of the objects PSR J1614-2230 and PSR J0348+0432 with $M=1.97\pm 0.04M_{\odot}$~\cite{Demorest} and $M=2.01\pm 0.04M_{\odot}$~\cite{Antoniadis}, lower and middle bands respectively. The upper band represents the new result of $2.14{{+0.10}\atop{-0.09}} M_{\odot}$ for the mass of the MSP J0740+6620 pulsar at $68.3\%$ credible level, recently presented in Ref.~\cite{Cromartie}. One can see that even with the overall reduction in the maximum mass that occurs in the presence of gluons, there is a range of $G_V$ that is consistent with the maximum mass observations. More precisely, in the absence of gluons, the range is $\frac{G_V}{G_S}> 0.02$, while with gluons it becomes $\frac{G_V}{G_S} > 0.21$ for the MSP J0740+6620 pulsar.  

The range of allowable parameters is further constrained by the recent mass-radius estimates extracted from the NICER data, namely, $M=1.44^{+0.15}_{-0.14}M_{\odot}$ with $R=13.02^{+1.24}_{-1.06}$~km~\cite{Col}, $M=1.34^{+0.15}_{-0.16}M_{\odot}$ with $R=12.71^{+1.14}_{-1.19}$~km~\cite{l21}, and $R_{1.44}>10.7$~km~\cite{l25}. These estimates are indicated by black dots in the figure with their corresponding error bars. Each dot then determines the corresponding range of allowable $G_V$.  In the case without gluons, for each dot there is a range of $G_V$ consistent with both constraints, from NICER's and the maximum mass. Adding the gluons reduces the compatibility to just one of NICER estimates, the one with $R_{1.44}>10.7$~km, which is the only that can overlap with the condition $\frac{G_V}{G_S}> 0.21$.

While the number of accurately measured masses is increasing steadily, the radii are much more 
difficult to obtain. The recent determination by the NICER group for the neutron star PSR J0030+0451 
is probably the most reliable measurement today. As pointed out, it predicts a radius about $11$~km 
to $13$~km for $M\sim 1.4M_\odot$. This range for $R$ is on the ``high'' side of expected values. 
Small radii reports have been presented over the years (see, for example, 
Refs.~\cite{bogdanov,ozelfreire}) although they involve some form of modeling and are not as direct. 
For example, the radius of the NS in the quiescent low-mass X-ray binary X5 has been constrained to 
$R=9.6^{+0.9}_{-1.1}$~km for a $M = 1.4 M_\odot$ NS, according to Ref.~\cite{bogdanov}. By 
considering this data instead of NICER's one, we would find that only the model with gluons, for 
$\frac{G_V}{G_S}\lesssim 0.1$, can reproduce it. It is clear that there are identified methods to 
infer the radii, and small values could ultimately be confirmed, but there is work to be done and 
questions on the road ahead that need to be answered~\cite{Lattimer}. Needless to say, this is a 
very important question because it may be indicative of a ``two family'' situation~\cite{alvarez} 
among other possibilities. 

Regarding results depicted in Fig.~\ref{mr}, we remark that CFL model with and without gluons 
predicts high values for the NS mass. In that direction, the detection of the unusual event 
GW190814~\cite{Abbot} featuring a member of the pair in the interval $(2.5 ‒ 2.67)M_{\odot}$ is 
important in the context of the maximum mass issue of NS's and the equation of state. Even though 
the object can well be a black hole (of the ``light'' type which has never been observed in the 
local Universe), there is mounting evidence that it could also be an extreme case of the compact 
star branch. This stems from i) the analysis of the LIGO-Virgo Collaboration showing that the 
``light'' object is an outlier from the BH distribution detected from merging, hence it should be on 
the compact star side~\cite{Abbottetal2021}; ii) the statistical evidence that the maximum mass 
$M_{max}$ is high, around $(2.5-2.6)M_\odot$~\cite{Alsing,Horvath} and iii) the studies that have 
argued the possible nature of the lighter object as a strange quark 
star~\cite{Bombaci,HorvathMoraes2021} for which the theoretical sequences can reach this higher 
level without obvious fatal problems. In summary, while we are not claiming that the GW190814 light 
component must be a  compact star, this possibility has been reinforced recently and guarantees 
extended studies, with clear connections with the subject of the present paper.

Now we need to consider the compatibility with the tidal deformability associated to the observation 
of GW emission from the binary star merger GW170817 event, detected by the LIGO/Virgo Collaboration 
(LVC)~\cite{Abbott2,lvc1,lvc2}. The GW emission caused an energy flux out of the binary system and 
produced the inspiral motion of the stars ~\cite{Taylor,Hulsel}. The obtained data allowed LVC to 
establish some constraints on $\Lambda_1$ and $\Lambda_2$. It was also possible to determine a range 
for $\Lambda_{1.4}$ (deformability of the star with $M=1.4M_\odot$). In order to calculate $\Lambda$ 
as a function of $M$ or $R$, Eq.~(\ref{diml}), one needs the second Love number $k_2$, which is 
defined as
\begin{align}
k_2 &=\frac{8C^5}{5}(1-2C)^2[2+2C(y_R-1)-y_R]\nonumber\\
&\times\Big\{2C [6-3y_R+3C(5y_R-8)] \nonumber\\
&+ 4C^3[13-11y_R+C(3y_R-2) + 2C^2(1+y_R)]\nonumber\\
&+ 3(1-2C)^2[2-y_R+2C(y_R-1)]{\rm ln}(1-2C)\Big\}^{-1},
\label{k2}
\end{align}
with $y_R\equiv y(R)$, and $y(r)$ obtained as the solution of 
\begin{align}
r\frac{dy}{dr} + y^2 + yF(r) + r^2Q(r) = 0,     
\label{dydr}
\end{align}
that has to be solved as part of a coupled system containing the TOV equations, Eqs.~(\ref{TOV1}) and~(\ref{TOV2}). Here, $F(r)$ and $Q(r)$ are defined as
\begin{eqnarray}
F(r) &=& \frac{1 - 4\pi r^2[\epsilon(r) - P(r)]}{f(r)}, 
\\
Q(r)&=&\frac{4\pi}{f(r)}\left[5\epsilon(r) + 9P(r) + 
\frac{\epsilon(r)+P(r)}{v_s^2(r)}- \frac{6}{4\pi r^2}\right]
\nonumber\\ 
&-& 4\left[ \frac{m(r)+4\pi r^3 P(r)}{r^2f(r)} \right]^2,
\label{qr}
\end{eqnarray}
where $v_s^2(r)=\partial P(r)/\partial\epsilon(r)$ is the squared sound velocity~\cite{tanj10,Prakash,hind08,damour,tayl09}. 

\begin{figure}[!htb]
\centering
\includegraphics[scale=0.35]{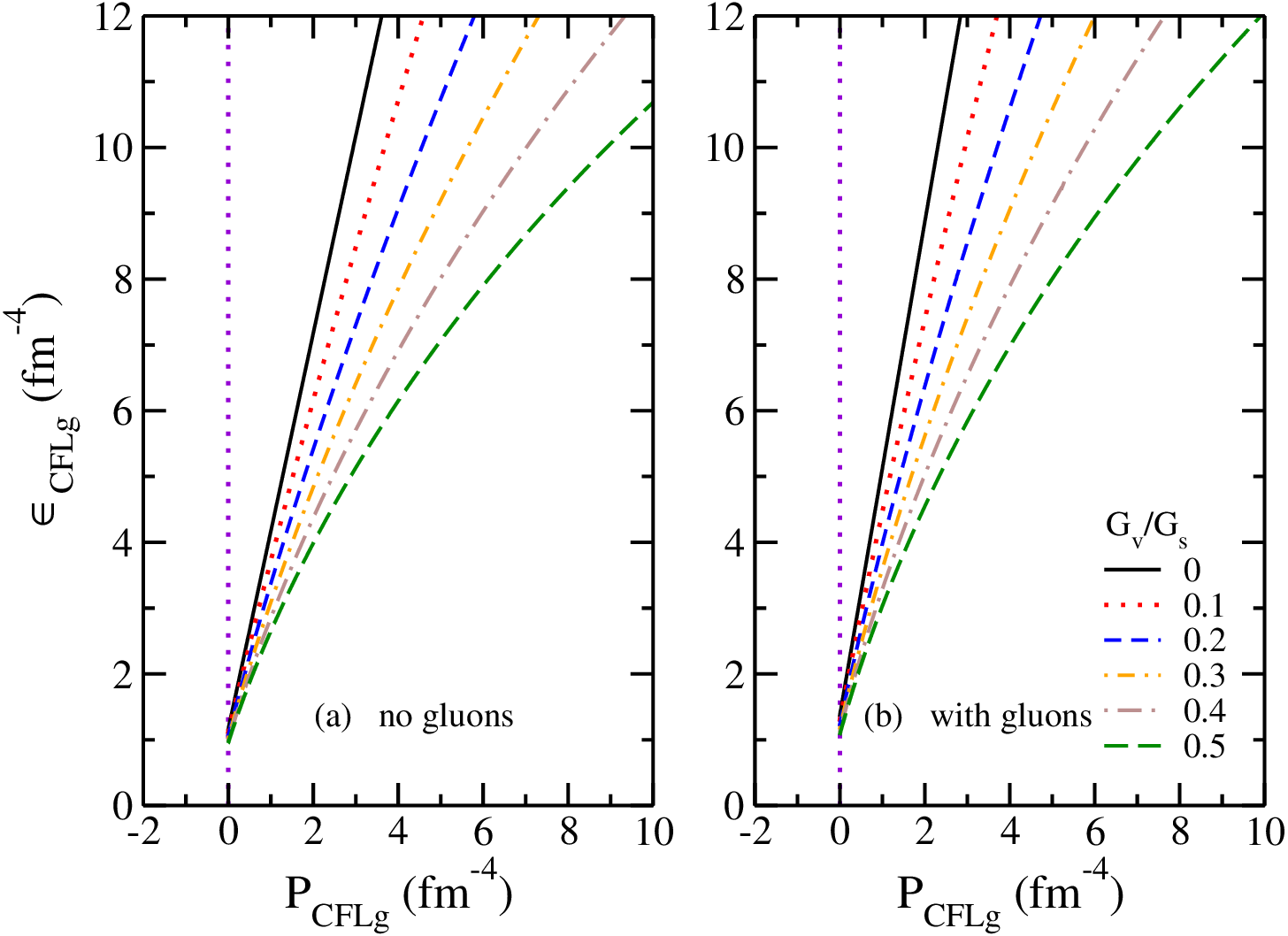}
\caption{Energy density as a function of pressure for the CFL phase (a) without and (b) with gluons contribution.}
\label{pe}
\end{figure}

While solving the TOV equations, the star surface is defined as the point where the pressure goes to zero, $P(R)=0$, as we mentioned before. Nevertheless, in the case of a bare strange star, the energy density is finite at this point as one can see in Fig.~\ref{pe}. 
This requires a correction to be added to the calculation of~$y_R$ to account for the energy discontinuity between the star's surface and its outside, reading~\cite{angli,wang,mingli,Takatsy2020}
\begin{equation}
y_R\rightarrow y_R - \frac{4\pi R^3\epsilon_s}{M},
\label{yr}
\end{equation}
where $\epsilon_s$ is the energy density difference between the internal and external regions.

\begin{figure}[!htb]
\centering
\includegraphics[scale=0.35]{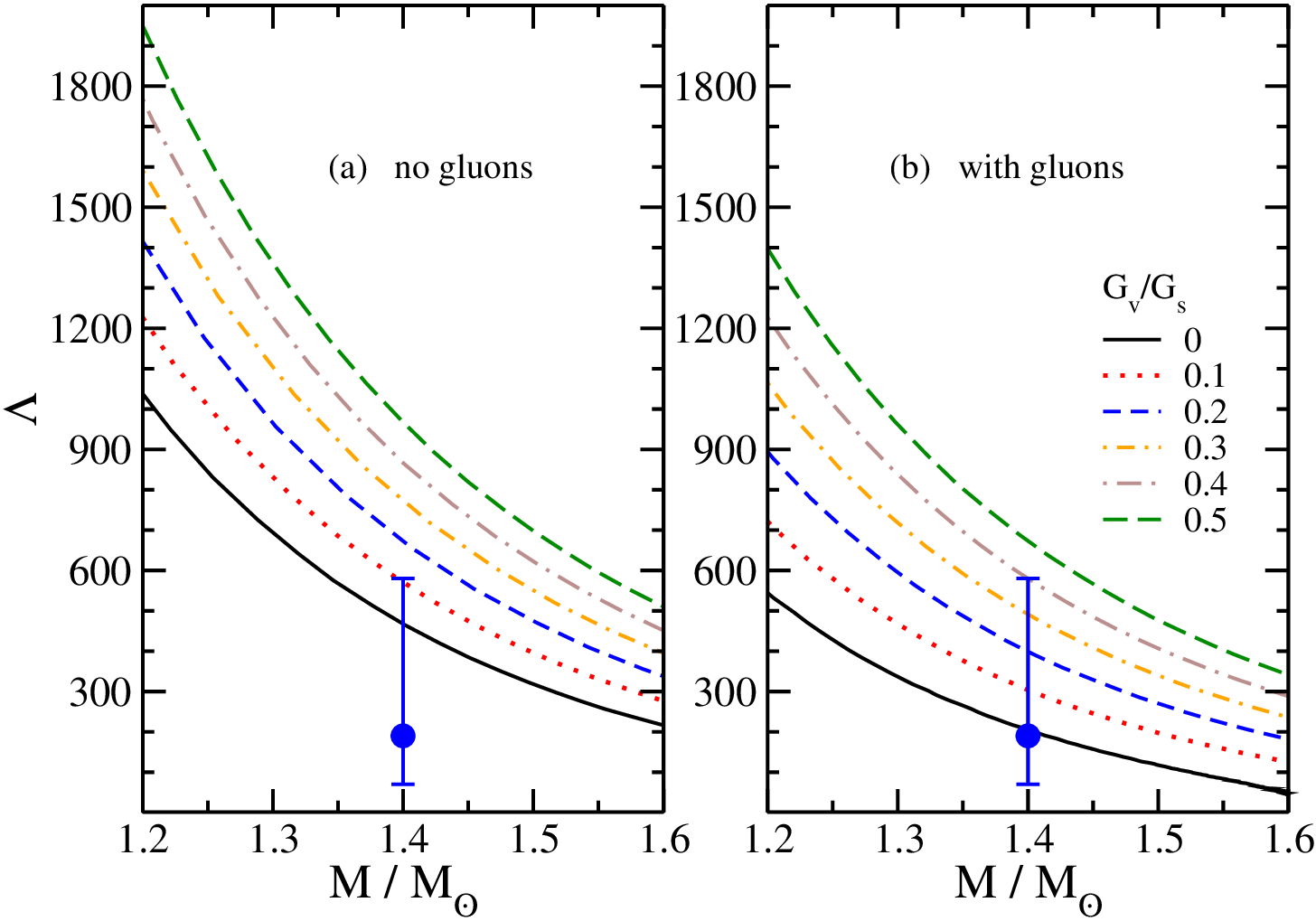}
\caption{Dimensionless tidal deformability as a function of quark star mass, in units of $M_\odot$, for the CFL phase (a) without and (b) with gluons contribution. Full circle: result of $\Lambda_{1.4}=190_{-120}^{+390}$ obtained by LVC~\cite{lvc2}.}
\label{lm}
\end{figure}

Since the TOV equations are solved coupled to Eq.~(\ref{dydr}) and Eq.~(\ref{yr}), it is possible to obtain the tidal deformabilities in the framework of the CFL model, with and without gluons, for different parametrizations generated by varying $G_V$. We compare these quantities with observational data extracted from LVC. In Fig.~\ref{lm}, we show the dimensionless tidal deformability as a function of $M$. 
From Fig. (\ref{lm}), one can gather that the vector interactions tend to increase $\Lambda$ at any given value of $M$ in both cases, i.e., with and without gluons. On the other hand, the effect of the gluons is to decrease the tidal deformability at any given $M$ and $G_V$.

\begin{figure}[!htb]
\centering
\includegraphics[scale=0.33]{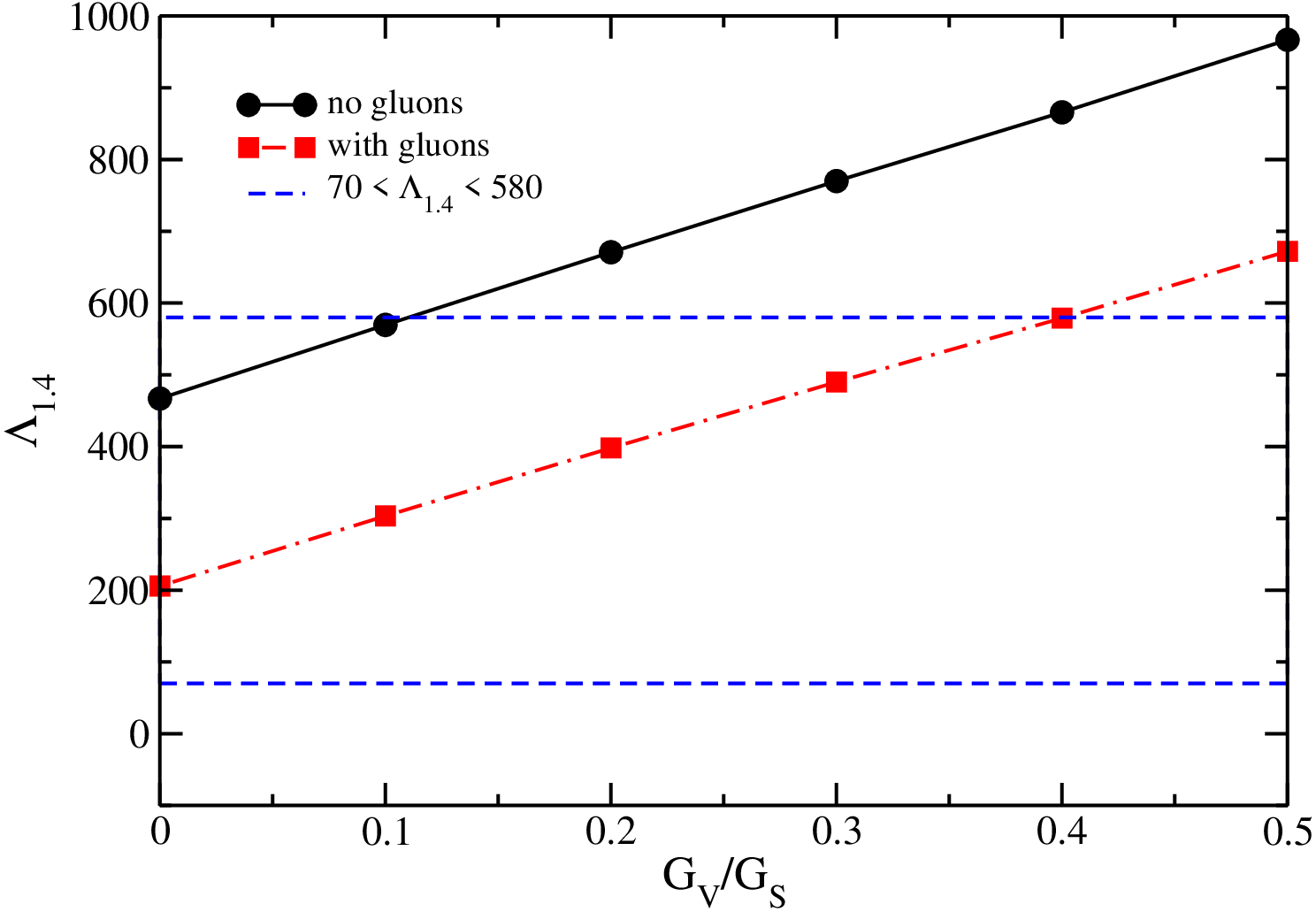}
\caption{$\Lambda_{1.4}$ as a function of the vector channel strength, in units of $G_S$, for the CFL phase with and without gluons contribution. Dashed blue lines: $\Lambda_{1.4}=190_{-120}^{+390}$ obtained by LVC~\cite{lvc2}.}
\label{l14-gv}
\end{figure}

For the specific case of $\Lambda_{1.4}$, in which one has an observational value determined from LVC, namely, $\Lambda_{1.4}=190_{-120}^{+390}$~\cite{lvc2} (GW170817 event), we see a clear trend in the CFL phase with gluon contribution to attain the LVC data. Furthermore, a clear linear increasing of $\Lambda_{1.4}$ as a function of $G_V$ is observed as displayed in Fig.~\ref{l14-gv} for both cases: with and without gluon contribution. In this figure, each circle/square represents a value of $\Lambda_{1.4}$ for each value of $G_V/G_S$. Notice that the parametrizations $0\leqslant G_V/G_S \leqslant 0.4$ are completely inside the GW170817 constraint of $\Lambda_{1.4}$ for the case with gluons. With no gluon contribution, this range becomes more stringent, namely, $0\leqslant G_V/G_S \leqslant 0.1$.

\begin{figure}[!htb]
\centering
\includegraphics[scale=0.35]{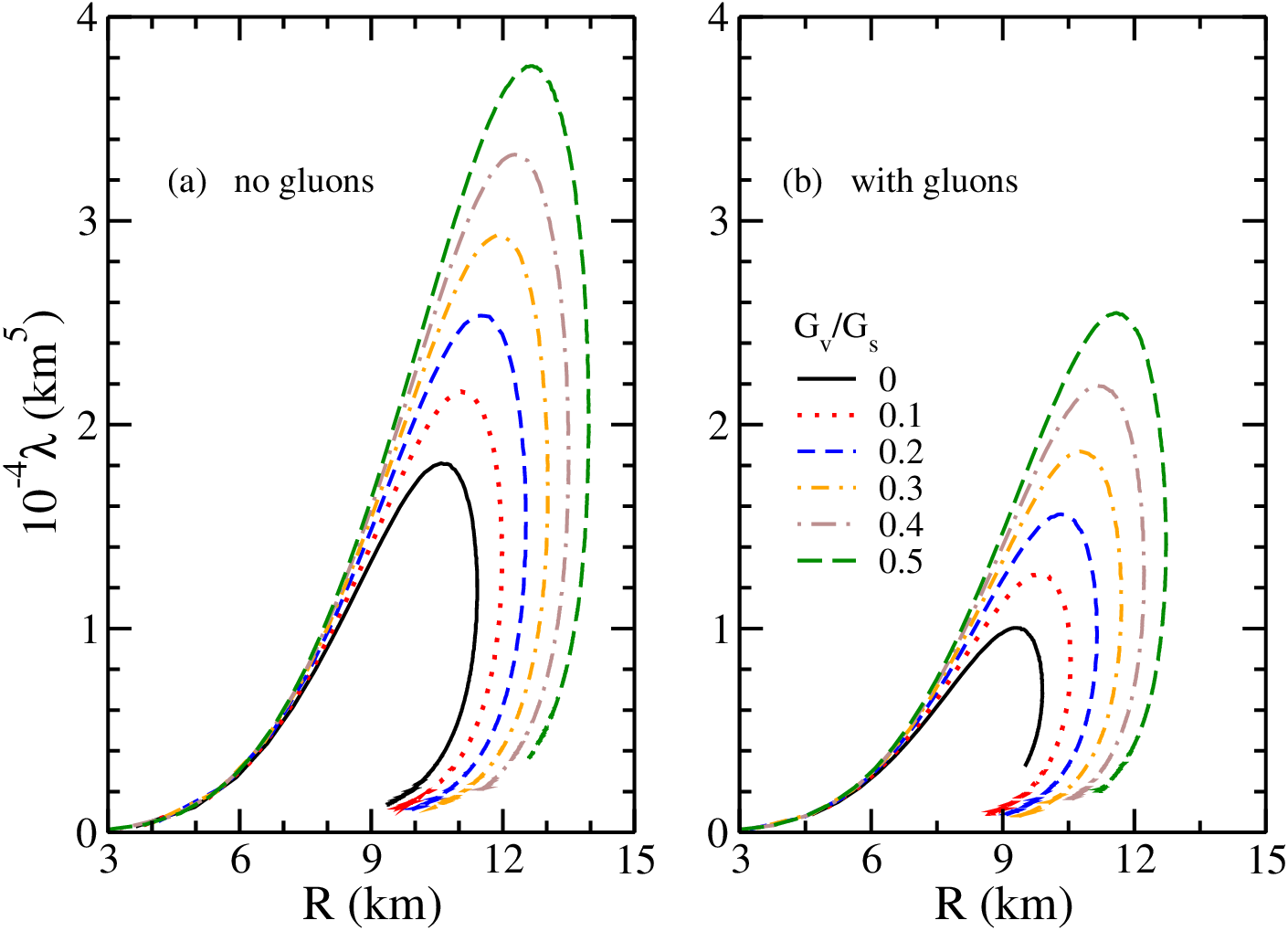}
\caption{$\lambda$ as a function of $R$ for the CFL phase (a) without and (b) with gluons contribution. }
\label{ldim}
\end{figure}

For the sake of completeness, we show in Fig.~\ref{ldim} how $\lambda$, calculated from Eq.~(\ref{lambda}), depends on the star radius~$R$.
We verify that the same features observed in Fig.~\ref{lm} are also presented in the $\lambda$ vs $R$ curves, namely, that $\lambda$ increases with $G_V$ at a fixed value of $R$, and that the gluon contribution reduces the $\lambda$ values. In addition, one can also see a reduction of the star radii for the model with gluons included. This is an effect also verified in the mass-radius profiles exhibited in Fig.~\ref{mr}.

\begin{figure}[!htb]
\centering
\includegraphics[scale=0.34]{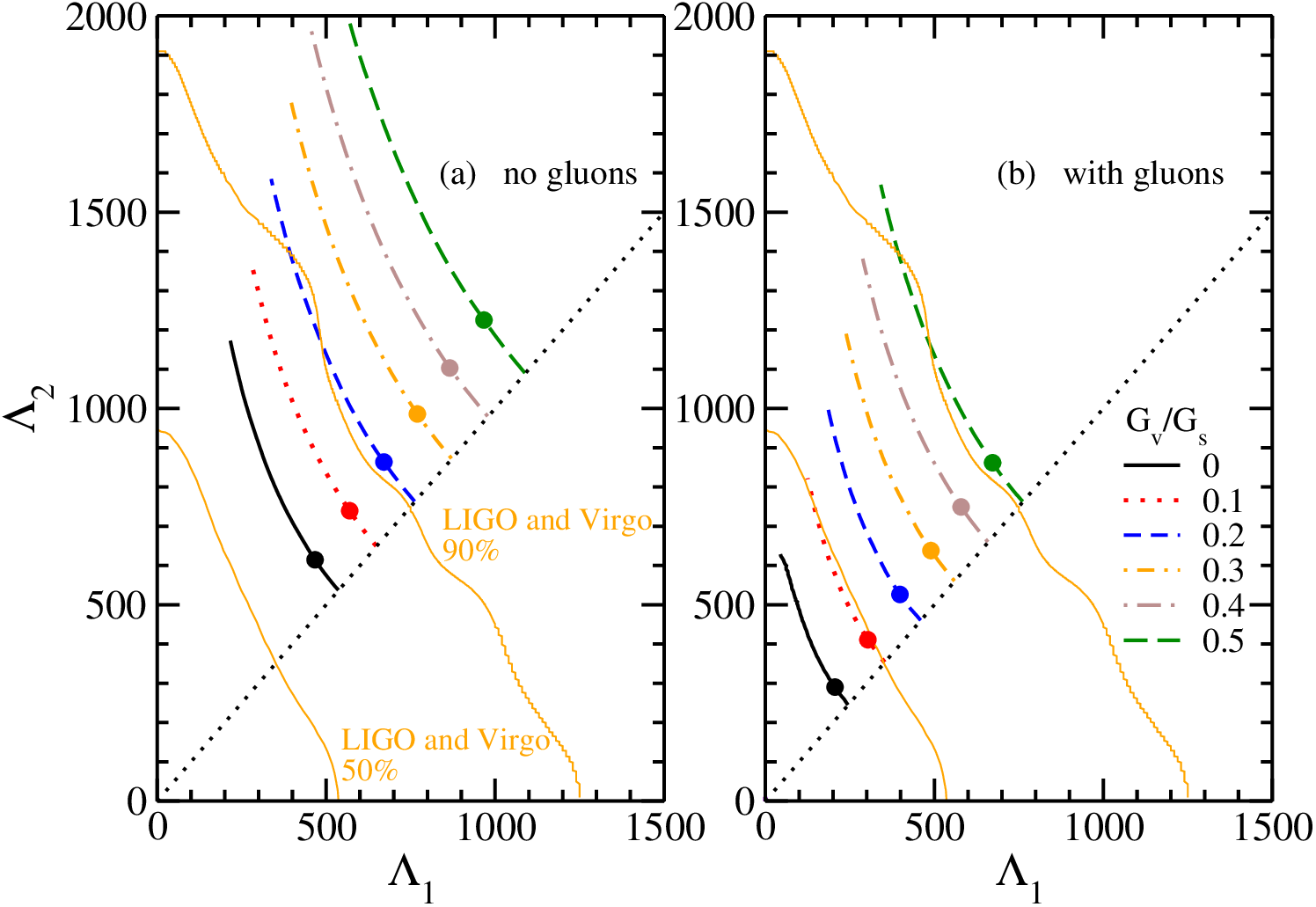}
\caption{Dimensionless tidal deformabilities for the case of high-mass ($\Lambda_1$) 
and low-mass ($\Lambda_2$) components of the GW170817 event for stars in the CFL phase (a) without and (b) with gluons contribution. The confidence lines (50\% and 90\%) are taken from Ref.~\cite{lvc2}. The dots on the curves denote the values that corresponds to $M_1=1.4M_\odot$ and $\Lambda_1=\Lambda_{1.4}$.}
\label{l1l2}
\end{figure}

In Fig.~\ref{l1l2} we show the tidal deformabilities $\Lambda_1$ and $\Lambda_2$ of the binary system in the CFL phase. We also depict the contour lines of $50\%$ and $90\%$ credible levels (full orange curves) related to the GW170817 event~\cite{lvc2}. 
In order to produce such curves, with different $G_V$ values, we run the mass of one of the stars, $M_1$, in the range of $1.37 \leqslant M_1/M_\odot \leqslant 1.60$~\cite{lvc2,Abbott2}. The mass of the second star, $M_2$, presents a relationship with $M_1$ via the chirp mass defined as \mbox{${\mathcal M} = (M_1M_2)^{3/5}/(M_1+M_2)^{1/5}$}~\cite{lvc1}. The analysis of the LVC provided ${\mathcal M}$ as presenting the value of $\mathcal{M}=1.188^{+0.004}_{-0.002}M_\odot$~\cite{lvc1}, that generates a variation of $1.17 \leqslant M_2/M_\odot \leqslant 1.36$~\cite{lvc1,lvc2} for the mass of the companion star. 

By comparing the curves in Fig.~\ref{l1l2}-$a$ and -$b$, we clearly notice (once more) that the inclusion of gluons contribution favors the curves to satisfy the LVC constraint in the $\Lambda_1\times\Lambda_2$ plane. In this case, all curves with $G_V\leqslant 0.4G_S$ are completely inside the $90\%$ credible region. The one in which $G_V=0.5G_S$ is in the limit of the external boundary curve. 

We also marked with a dot the points in the curves where $M_1=1.4M_\odot$ and, consequently, $\Lambda_1=\Lambda_{1.4}$. From such points we can observe a connection between the results shown in Figs.~\ref{l14-gv} and~\ref{l1l2}. The decreasing of $G_V$ implies lower values of $\Lambda_{1.4}$, and in the case of the CFL phase studied here, it leads to an agreement with the LVC constraint for this quantity, as pointed out before. The same kind of compatibility is verified in the entire $\Lambda_1\times\Lambda_2$ curves. The reduction of $\Lambda_{1.4}$, due to the decreasing of $G_V$, is followed by a shift of all the curves toward the observational region predicted by the GW170817 event. This is a feature observed for the CFL phase including or not the gluon contribution. We also remark here that the magnitude of the curves exhibited in Figs.~\ref{lm} and~\ref{l1l2} are compatible with those obtained by relativistic and nonrelativistic hadronic models~\cite{had1}, which are also in agreement with the observational data reported by LVC, see for example Refs.~\cite{had2,had3,had4,had5}. 
\begin{figure}[!htb]
\centering
\includegraphics[scale=0.34]{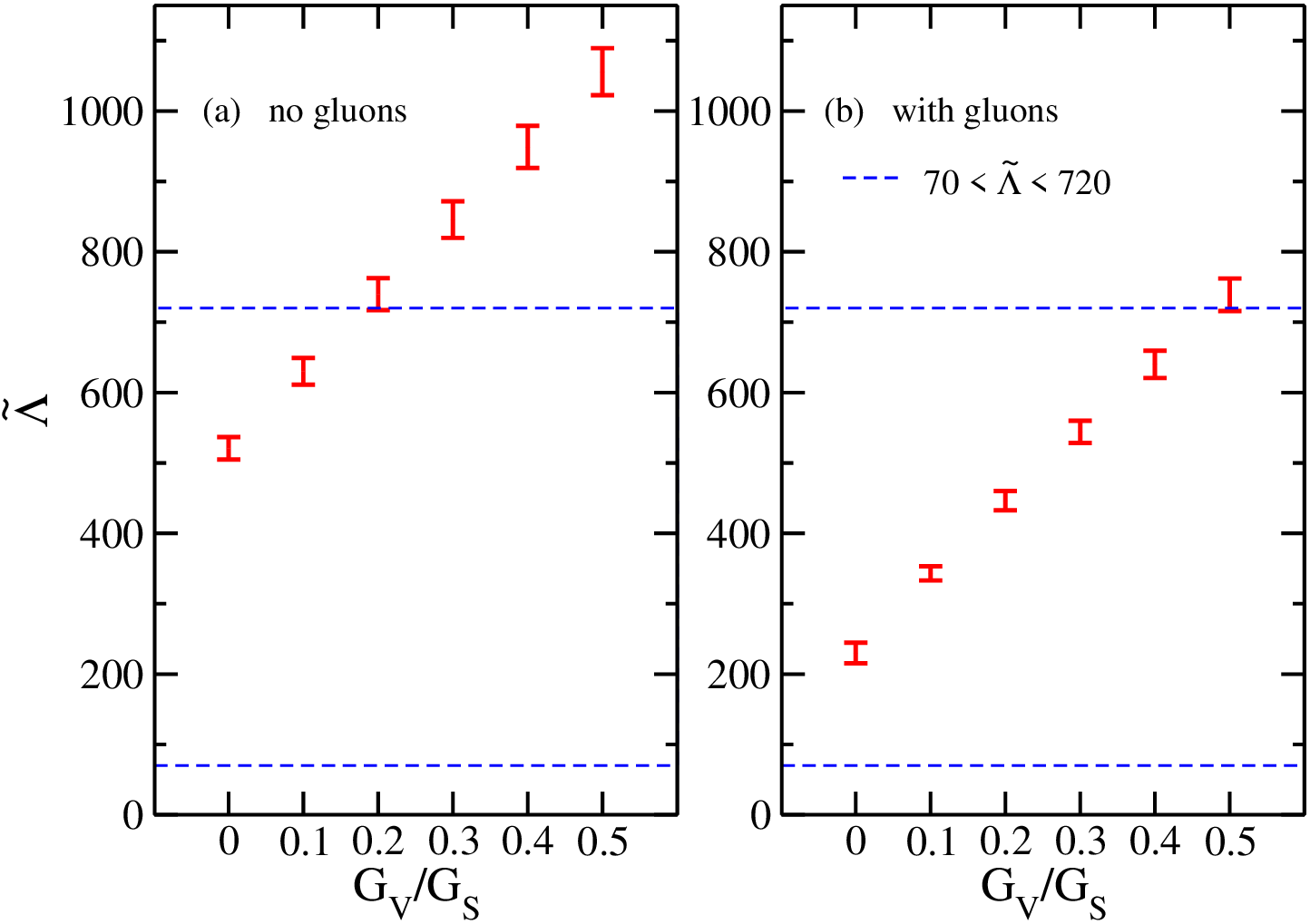}
\caption{$\tilde{\Lambda}$ for different parametrizations (different $G_V$ values) for strange stars in the CFL phase (a) without and (b) with gluons contribution. Dashed lines: range of $\tilde{\Lambda}=300^{+420}_{-230}$ determined by LVC~\cite{Abbott2}.}
\label{ltilde}
\end{figure}

Finally, Fig.~\ref{ltilde} shows the ranges of $\tilde{\Lambda}$, Eq.~(\ref{tilde}), obtained for strange stars in the CFL phase. $\tilde{\Lambda}$ is calculated as a function of the mass of one of the stars forming the binary system, namely, $\tilde{\Lambda}=\tilde{\Lambda}(M_1)$ or $\tilde{\Lambda}=\tilde{\Lambda}(M_2)$. Since $M_1$ (or $M_2$) is defined into a particular range according to the GW170817 event, each parametrization with a fixed $G_V$ value will produce a range for $\tilde{\Lambda}$. We compare the results with the constraint on the combined dimensionless tidal deformability obtained by LVC, namely, $\tilde{\Lambda}=300^{+420}_{-230}$~\cite{Abbott2}. Once again, the CFL phase with the gluon contribution supports the observational data from the GW170817 event. Just as with the behavior between $\Lambda_{1.4}$ and $G_V$ depicted in Fig.~\ref{l14-gv}, there is also a strong linear relation between $\tilde{\Lambda}$ and $G_V$. Lastly, the parametrizations $G_V/G_S < 0.2$~($G_V/G_S < 0.5$) for the CFL phase without~(with) gluons satisfy the constraint imposed by the observational range of $\tilde{\Lambda}$.

\section{Summary and concluding remarks}
\label{conclusions}

In this paper, we explored the compatibility of strange stars in the CFL phase with a set of observational constraints obtained from the GW170817 event, the maximum stellar mass from PSR J1614-2230~\cite{Demorest}, PSR J0348+0432~\cite{Antoniadis}, and MSP J0740+6620~\cite{Cromartie}; and the mass-radius estimates from recent NICER data. An important goal of this paper has been to present a systematic approach to test the observational compatibility of a quark star in a particular phase.

We considered an absolutely stable strange star made of massless $u$, $d$, and $s$ quarks in the CFL phase modeled by a NJL theory with diquark and vector interaction channels. Gluon effects were incorporated by adding the gluon self-energy calculated in the finite-density color superconducting medium \cite{Our-Gluons} to the thermodynamic potential.

\begin{figure}[!t]
\centering
\includegraphics[scale=0.34]{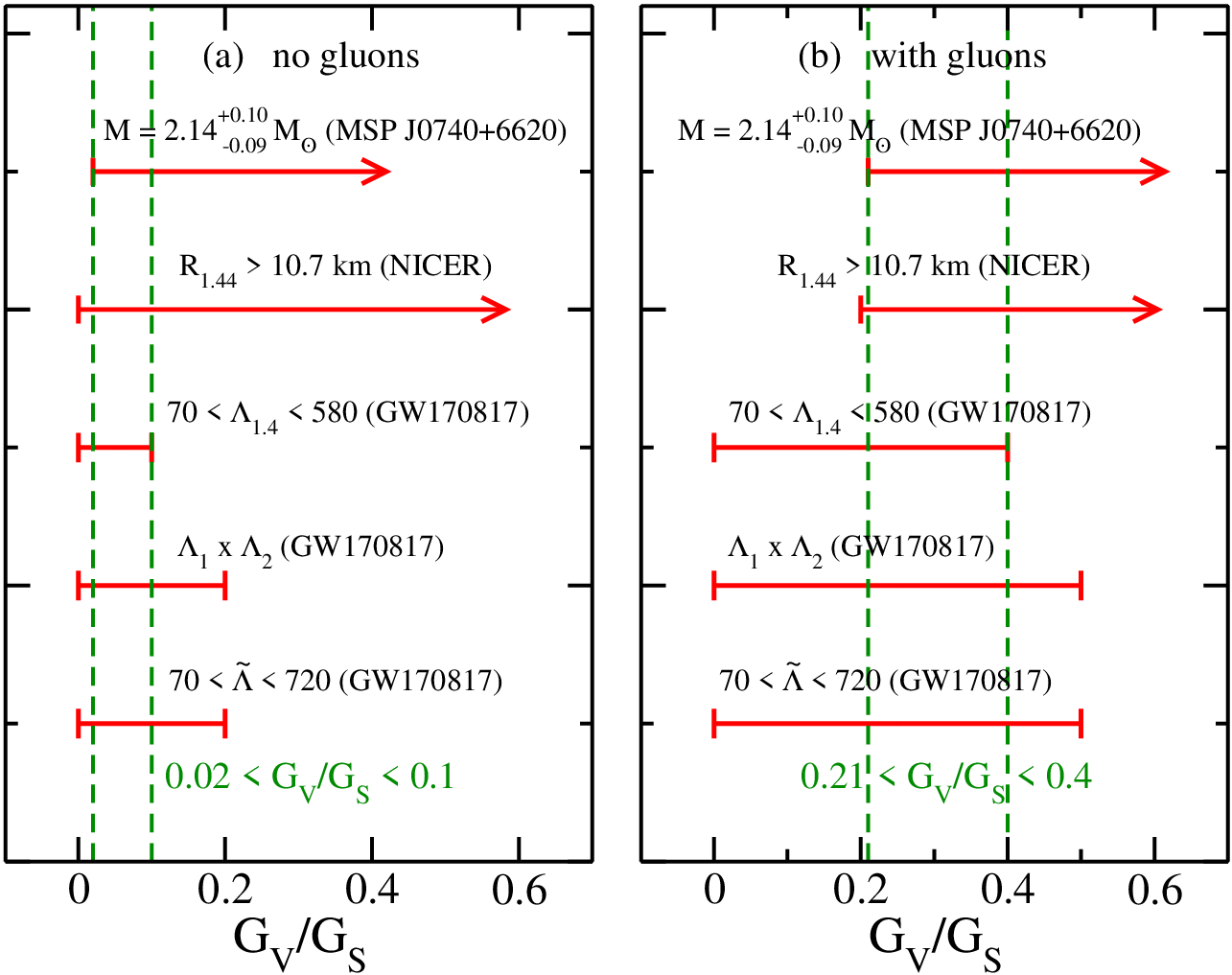}
\caption{Compilation of $G_V$ ranges predicted by this work from comparison with astrophysical data. The intersection of the ranges is (a) $0.02<G_V/G_S<0.1$, and (b) $0.21<G_V/G_S<0.4$ for the model without and with gluons contribution included, respectively.}
\label{finalranges}
\end{figure}

In Fig.~\ref{finalranges} we summarize our main findings and the range of overall compatibility for the CFL phase, with and without the gluon term. The  regions between the dashed vertical lines in Figs.~\ref{finalranges}{\color{blue}a} and~\ref{finalranges}{\color{blue}b} indicate the range of $G_V/G_S$ compatible with all the constraints simultaneously. In general, including gluons tends to better accommodate the tidal deformability observations. Gluons also contribute  to widen the range of vector interactions compatible with all the observations. At the same time, they increase the minimum $G_V$ needed to satisfy the constraints, although the resultant range is still within the theoretically acceptable values of vector interaction strengths. 

Our results show that the CFL phase, with or without the gluon contribution, is compatible with the set of recent observations considered in this paper. This of course does not ensure that future observations and/or updated refinement of the estimated values from known observations cannot push the CFL phase out of the compatibility region. Even in such a case, other phases that can be realized in a strange star would be worth to be examined against the new constraints using the same approach followed here. 

It is interesting that the dimensionless tidal deformabilities of CFL stars found in this paper are comparable to those of hadronic stars \cite{had2,had3,had4,had5}, i.e., they have the same order of magnitude within the allowable parameter range.

Finally, we call the reader's attention to the fact that strictly speaking, the CFL phase of massless $u$, $d$ and~$s$ quarks is energetically favored only at asymptotically large densities (i.e. at densities much higher than the $s$ quark mass). At more realistic densities, the effect of the $s$ quark mass may lead to chromomagnetic instabilities and eventually to an spatially inhomogenous phase \cite{chromomagnetic inst}. At those densities, other phases may compete with the CFL phase and become plausible candidates for the strange star phase. Along this direction, inhomogeneous phases of dense quark matter with chiral quark-hole condensates have been attracting much interest in recent years \cite{inh-con}. In this context, one of those phases, the so called magnetic dual chiral density wave (MDCDW) phase, has emerged as a viable candidate, which so far has satisfied some important astrophysical constraints, as for instance the observed $\sim 2M_{\odot}$~\cite{Carignano}, and more recently its stability against collective fluctuations \cite{MDCDW stability}, ensuring its robustness at the density and temperature conditions of neutron stars.

\begin{acknowledgments}
The work of E.J.F. and V.I. was supported in part by NSF grant PHY-2013222. The work of C.H.L., M.D., and O.L. is part of the project INCT-FNA proc. No. 464898/2014-5. It is also supported by Conselho Nacional de Desenvolvimento Científico e Tecnológico (CNPq) under Grants No. 310242/2017-7, 312410/2020-4, 406958/2018-1 (O.L.), and No. 433369/2018-3 (M.D.). We also acknowledge Funda\c{c}\~ao de Amparo \`a Pesquisa do Estado de S\~ao Paulo (FAPESP) under Thematic Project 2017/05660-0 (O.L., M.D., C.H.L), Grant No. 2020/05238-9 (O.L., M.D., C.H.L), and Thematic Project 2013/26258-4 (L.P.). J.E. Horvath has been supported by Fapesp Agency (S\~ao Paulo) and CNPq
(Federal Agency, Brazil) through grants and scolarships.
\end{acknowledgments}


\begin{thebibliography}{}

\bibitem{Abbott2017}
 B.P. Abbott, R. Abbott, T.D.Abbott, F. Acernese, et al., Astrophys. J. Lett {\bf 848}, L12 (2017).
 
 \bibitem{AbbottGRB}
 B.P. Abbott, R. Abbott, T.D.Abbott, F. Acernese, et al., Astrophys. J. Lett {\bf 848}, L13 (2017).
 
\bibitem{Tsvi}
 O. Gottlieb, E. Nakar, T. Piran and K. Hotokezaka, Mon. Not. R. Astron. Soc. {\bf 479}, 588 (2018).
 
 \bibitem{Istvan}
 I. Horv\'ath, B.G. T\'oth, J. Hakkila, et al., Astrophys. Space Sci. {\bf 363}, 53 (2018).

 \bibitem{Valenti}
 S. Valenti  et al., Astrophys. J. Lett {\bf 848}, L24 (2017).

\bibitem{Nucleo}
D.M. Siegel, Eur. Phys. J. A {\bf 55}, 203 (2019).

\bibitem{Nucleo2}
B. Metzger, Living Reviews in Relativity {\bf 20}, 3 (2017).

\bibitem{Watson2019}
Watson, D., Hansen, C.J., Selsing, J. et al., Nature {\bf 574}, 497 (2019).

\bibitem{Nucleo3}
D. Kasen et al., Nature {\bf 551}, 80 (2017).

\bibitem{ref1} N. K. Glendenning, S. A. Moszkowski, Phys. Rev. Lett., {\bf 67}, 2414 (1991).

\bibitem{ref2} I. Bombaci, P. K. Panda, C. Providencia, I. Vida\~na, Phys. Rev. D {\bf 77}, 083002 (2008).

\bibitem{ref3} V. Dexheimer, S. Schramm, Astrohys. J. {\bf 683}, 943 (2008).

\bibitem{ref4} G. F. Burgio, H.-J. Schulze, A. Li, Phys. Rev. C {\bf 83}, 025804 (2011).

\bibitem{ref5} I. Vida\~na, D. Logoteta, C. Providencia, I. Bombaci, Europhys. Lett. {\bf 94}, 11002 (2011).

\bibitem{ref6} L. Bonanno, A. Sedrakian, Astron. Astrophys. {\bf 539}, A16 (2012).

\bibitem{ref7} I. Bednarek, P. Haensel, J. L. Zdunik, M. Bejger and R. Manka, Astron. Astrophys. {\bf 543}, A157 (2012).

\bibitem{Yo} J. E. Horvath and R. A. de Souza, J. Phys. Conf. Ser. {\bf 861}, 012010 (2016). 

\bibitem{Bodmer}
A. R. Bodmer, Phys. Rev. D {\bf 04}, 1601 (1971).

\bibitem{Witten}
E. Witten, Phys. Rev. D {\bf 30}, 272 (1984).

\bibitem{Terazawa} H. Terazawa, K. Akama and Y. Chikashige, Prog. Theor. Phys. {\bf 60}, 1521 (1978).

\bibitem{Itoh}
N. Itoh, Prog. Theor. Phys. {\bf 44}, 291 (1970).

\bibitem{SS-Evidences} I. Bombaci, Phys. Rev. C {\bf 55}, 1587 (1997); K.S. Cheng, Z.G. Dai, D.M. Wai, and T. Lu, Science {\bf 280}, 407 (1998); X.–D. Li, I. Bombaci, M. Dey, J. Dey, and E.P.J. van den Heuvel, Phys. Rev. Lett. {\bf 83}, 3776 (1999); X.–D. Li, S. Ray, J. Dey, M. Dey, and I. Bombaci, Astrophys. J. {\bf 527}, L51 (1999); E. J. Ferrer and V. de la Incera, arXiv: 2010.02314.

\bibitem{nos}
J.E. Horvath {\it et al.}, Universe {\bf 5}, 144 (2019).

\bibitem{lvc1} B. P. Abbott {\it et al}. (LIGO Scientific Collaboration and Virgo Collaboration), Phys. Rev. Lett. {\bf 119}, 161101 (2017).

\bibitem{lvc2} B. P. Abbott {\it et al}., (LIGO Scientific Collaboration and Virgo Collaboration), Phys. Rev. Lett. {\bf 121}, 161101 (2018).

\bibitem{Abbott2} B. P. Abbott et al., Phys. Rev. X {\bf 9}, 011001 (2019).

\bibitem{Our-Gluons} E. J. Ferrer, V. de la Incera and L. Paulucci, Phys. Rev. D {\bf 92}, 043010 (2015).

\bibitem{Demorest} P.B. Demorest,T. Pennucci, S.M. Ransom, M.S.E. Roberts and J.W.T.  Hessels, Nature {\bf 467}, 1081 (2010).
 
\bibitem{Antoniadis} J. Antoniadis, P.C.C. Freire, N.  Wex, T. Tauris {\it  et al.}, Science {\bf 340},  6131 (2013).
 
\bibitem{Cromartie} H. T. Cromartie, E. Fonseca, S. M. Ransom, P. Demorest {\it et al}, Nature Astronomy {\bf 4}, 72 (2020).

\bibitem{had2} O. Louren\c{c}o, M. Dutra, C. H. Lenzi, C. V. Flores, and D. P.
Menezes, Phys. Rev. C {\bf 99}, 045202 (2019). 

\bibitem{had3} O. Louren\c{c}o, M. Dutra, C. H. Lenzi, M. Bhuyan, S. K. Biswal, and B. M. Santos, Astrophys. J. {\bf 882}, 67 (2019).

\bibitem{had4} O. Louren\c{c}o, M. Dutra, C. H. Lenzi, S. K. Biswal, M. Bhuyan, and D. P. Menezes, Eur. Phys. J. A {\bf 56}, 32 (2020).

\bibitem{had5} O. Louren\c{c}o, M. Bhuyan, C. H. Lenzi, M. Dutra, C. Gonzalez-Boquera, M. Centelles, and X. Vi\~nas, Phys. Lett. B {\bf 803}, 135306 (2020).
 
\bibitem{CS}
B. C. Barrois, {\it Nucl. Phys. B} {\bf129} 390 (1977); S. Frautschi, in: Proceedings of the Workshop on Hadronic Matter at Extreme Energy Density, N. Cabibbo, ed (Erice, Italy 1978); D. Bailin and A. Love, Phys. Rep. {\bf 107}, 325 (1984).

\bibitem{CFL}
M. Alford, K. Rajagopal and F. Wilczek, Phys. Lett. B {\bf 422}, 247 (1998).

\bibitem{DiQuark} L. Paulucci , E.J. Ferrer, J.E. Horvath, V. de la Incera, J. Phys. G {\bf 40}, 125202 (2013); E. J. Ferrer, V. de la Incera, J. P. Keith, I. Portillo, Nucl. Phys. A {\bf 933}, 229 (2015).

\bibitem{Vector-Int} 
M. Kitazawa, T. Koide, T, Kunihiro and Y. Nemoto, Prog. Theor. Phys. {\bf 108}, 929 (2002).

\bibitem{dynamical} T. E. Restrepo, J. C. Macias, M. B. Pinto, and G. N. Ferrari, Phys. Rev. D {\bf 91}, 065017 (2015).

\bibitem{Gluon-Mass}
D. T. Son and M. A. Stephanov, Phys. Rev. D {\bf 61}, 074012 (2000); M. Rho, E. Shuryak, A. Wirzba, and I. Zahed, Nucl. Phys. A {\bf 676}, 273 (2000); S. R. Beane, P. F. Bedaque, and M. J. Savage, Phys. Lett. B {\bf 483}, 131 (2000); K. Zarembo, Phys. Rev. D {\bf 62}, 054003 (2000); D. H. Rischke, Phys. Rev. D {\bf 62}, 054017 (2000).

\bibitem{Israel}E. J. Ferrer, V. de la Incera, J. P. Keith, I. Portillo, P. L. Springsteen, Phys. Rev. C {\bf 82}, 065802 (2010).


\bibitem{Oertel} M. Buballa and M. Oertel, Phys. Lett. B {\bf 457}, 261 (1999).

\bibitem{kata} K. Chatziioanou, Gen. Rel. Grav. {\bf 52}, 109 (2020).

\bibitem{Recons} F. Morawski and M. Bejger, Astron. Astrophys. {\bf 642}, A78 (2020).

\bibitem{Col} M. C. Miller {\it et al}., Astrophys. J. Lett. {\bf 887}, L24 (2019).

\bibitem{FH} E. E. Flanagan and T. Hinderer, Phys. Rev. D {\bf 77}, 021502(R) (2008).

\bibitem{Rehberg}P. Rehberg, S. P. Klevansky and J. H\"ufner, Phys. Rev. C {\bf 53}, 410 (1996).

\bibitem{GD-GS} G. Pagliara and J.Schaffner-Bielich, Phys. Rev. D {\bf 77} 063004 (2008).

\bibitem{GV-Vacuum} T. Kunihiro, Phys. Lett. B {\bf 271} 395, (1991); R. Rapp, T. Schafer, E. V. Shuryak and M. Velkovsky, Phys. Rev. Lett. {\bf 81}, 53 (1998); K. Kashiwa, M. Matsuzaki, H. Kouno and M. Yahiro, Phys. Lett. B {\bf 657}, 143 (2007); J. Steinheimer and S. Schramm, Phys. Lett. B {\bf 736}, 241 (2014).

\bibitem{l21} T. E. Riley, A. L. Watts, S. Bogdanov, P. S. Ray, R. M. Ludlam, S. Guillot {\it et al}., Astrophys. J. Lett. {\bf 887}, L21 (2019).

\bibitem{l25} Slavko Bogdanov, Sebastien Guillot, Paul S. Ray, Michael T. Wolff, Deepto Chakrabarty, Wynn C. G. Ho {\it et al}., Astrophy. J. Lett. {\bf 887}, L25 (2019).

\bibitem{bogdanov} Slavko Bogdanov, Craig O. Heinke, Feryal \"Ozel, and Tolga G\"uver, Astrophys. J. 
{\bf 831}, 184 (2016).

\bibitem{ozelfreire} F. Ozel and P. Freire, Annu. Rev. Astron. Astrophys. {\bf 54}, 401 (2017).

\bibitem{Lattimer} J. Lattimer, AIP Conf. Proc. {\bf 2127}, 020001 (2019).

\bibitem{alvarez} D.E. Alvarez-Castillo and D. B. Blaschke, Phys. Rev. C {\bf 96}, 045809 (2017).

\bibitem{Abbot} R. Abbott {\it et al}., Astrophys. J. Lett. {\bf 896}, L44 (2020).

\bibitem{Abbottetal2021} R. Abbott {\it et al}., arXiv:2010.14533.

\bibitem{Alsing} J. Alsing, H. O. Silva and E. Berti, Mon. Not. R. Astron. Soc. {\bf 478}, 1377 (2018).

\bibitem{Horvath} J. E. Horvath {\it et al}., submitted (2021).

\bibitem{Bombaci} I. Bombaci {\it et al}., arXiv:2010.01509.

\bibitem{HorvathMoraes2021} J. E. Horvath and P. H. R. S. Moraes, Int. Jour. Mod. Phys. D {\bf 30}, 2150016 (2021).
 

\bibitem{Hulsel} R. A. Hulse and J. H. Taylor, Astrophys. J. {\bf 191} L59 (1974); {\bf 195} L51 (1975).

\bibitem{Taylor} J. H. Taylor, L. A. Fowler and P. M. McCulloch, Nature {\bf 277}, 437 (1979).

\bibitem{tanj10} T. Hinderer, B. D. Lackey, Ryan N. Lang, J. S. Read, Phys. Rev. D {\bf 81}, 123016 (2010).

\bibitem{hind08} T. Hinderer, Astrophys. J. {\bf 677}, 1216 (2008).

\bibitem{damour}  T. Damour, A. Nagar, Phys. Rev. D {\bf 81}, 084016 (2010).

\bibitem{tayl09} T. Binnington, E. Poisson, Phys. Rev. D {\bf 80}, 084018 (2009).

 \bibitem{Prakash} S. Postnikov, M. Prakash and J. M. Lattimer, Phys. Rev. D {\bf 82}, 024016 (2010). 

\bibitem{angli} En-Ping Zhou, Xia Zhou, and Ang Li, Phys. Rev. D {\bf 97}, 083015 (2018).

\bibitem{wang} Qingwu Wang, Chao Shi, and Hong-Shi Zong, Phys. Rev. D {\bf 100}, 123003 (2019).

\bibitem{mingli} Cheng-Ming Li, Shu-Yu Zuo, Yan Yan, Ya-Peng Zhao, Fei Wang, Yong-Feng Huang, and Hong-Shi Zong, Phys. Rev. D {\bf 101}, 063023 (2020).

\bibitem{Takatsy2020} J. Takatsy and P. Kovacs, Phys. Rev. D {\bf 102}, 028501 (2020).

\bibitem{had1} O. Louren\c{c}o, M. Dutra, and D. P. Menezes, Phys. Rev. C {\bf 95}, 065212 (2017); O. Louren\c{c}o, B. M. Santos, M. Dutra, and A. Delfino, Phys. Rev. C {\bf 94}, 045207 (2016).

\bibitem{chromomagnetic inst} M. Alford, J Berges, and K. Rajagopal, Nucl. Phys. B {\bf 558}, 219 (1999).

\bibitem{inh-con} I. E. Frolov, V. Ch. Zhukovsky and K. G. Klimenko, Phys. Rev. D {\bf 82}, 076002 (2010); T. Tatsumi, K. Nishiyama and S. Karasawa, Phys. Lett. B {\bf 743}, 66 (2015); E. J. Ferrer and V. de la Incera, Phys. Lett. B 769 (2017) 208; Nucl. Phys. B {\bf 931}, 192 (2018); Phys. Rev. D {\bf 102}, 014010 (2020).

\bibitem{Carignano} S. Carignano, E. J. Ferrer, V. de la Incera, and L. Paulucci, Phys. Rev. D {\bf 92}, 105018 (2015).

\bibitem{MDCDW stability} E. J. Ferrer and V. de la Incera, Phys. Rev. D {\bf 102}, 014010 (2020).

\end{thebibliography}
\end{document}